PLoS one

# Presymptomatic Risk Assessment for Chronic Non-Communicable Diseases

Badri Padhukasahasram[1]*[9][¤a], Eran Halperin[1][9][¤b][¤c], Jennifer Wessel[1][¤d], Daryl J. Thomas[1][¤e], Elana Silver[1], Heather Trumbower[1], Michele Cargill[1][¤f], Dietrich A. Stephan[1,2,3,4]

1 Navigenics, Foster City, California, United States of America, 2 Institute for Individualized Health, Palo Alto, California, United States of America, 3 The Cancer Genome Institute, Philadelphia, Pennsylvania, United States of America, 4 Children's Hospital Informatics Program, Children's Hospital Boston, Boston, Massachusetts, United States of America

## Abstract

The prevalence of common chronic non-communicable diseases (CNCDs) far overshadows the prevalence of both monogenic and infectious diseases combined. All CNCDs, also called complex genetic diseases, have a heritable genetic component that can be used for pre-symptomatic risk assessment. Common single nucleotide polymorphisms (SNPs) that tag risk haplotypes across the genome currently account for a non-trivial portion of the germ-line genetic risk and we will likely continue to identify the remaining missing heritability in the form of rare variants, copy number variants and epigenetic modifications. Here, we describe a novel measure for calculating the lifetime risk of a disease, called the genetic composite index (GCI), and demonstrate its predictive value as a clinical classifier. The GCI only considers summary statistics of the effects of genetic variation and hence does not require the results of large-scale studies simultaneously assessing multiple risk factors. Combining GCI scores with environmental risk information provides an additional tool for clinical decision-making. The GCI can be populated with heritable risk information of any type, and thus represents a framework for CNCD pre-symptomatic risk assessment that can be populated as additional risk information is identified through next-generation technologies.

Funding: This project was funded by Navigenics Inc. only in that all the authors were employees of Navigenics when this study was carried out. No other company funded this study. The study was carried out at Navigenics Inc. and the methodology used in this paper was initially developed for genetic risk calculations for customers using their genotype data in the product that is offered by Navigenics Inc. The research in this article was conducted to validate this methodology and its assumptions using real and simulated data. The funders had a role in study design, data collection and in the decision to publish this manuscript.

Competing Interests: All the authors were employed at Navigenics Inc when this study was carried out. No other company mentioned in the author affiliations was involved in the study. This is a primary research article and the data used in this study (WTCCC data) was not generated by the company and is available to all qualified researchers. This does not alter the authors' adherence to all the PLoS ONE policies on sharing data and materials.

* E-mail: bpadhuka@ucdavis.edu

9 These authors contributed equally to this work.

¤a Current address: Section on Ecology and Evolution and Genome Center, University of California Davis, Davis, California, United States of America
¤b Current address: Blavatnik School of Computer Science, Department of Molecular Microbiology and Biotechnology, Tel-Aviv, University, Tel-Aviv, Israel
¤c Current address: International Computer Science Institute, Berkeley, California, United States of America
¤d Current address: School of Medicine, Indiana University, Indianapolis, Indiana, United States of America
¤e Current address: Life Technologies, Foster City, California, United States of America
¤f Current address: Locus Development, San Francisco, California, United States America

## Introduction

Common chronic non-communicable diseases (CNCDs) are caused by a combination of genetic and environmental risk factors. These diseases account for the majority of disease burden, and the majority of health care cost, globally. Pre-symptomatic risk assessment of an individual for CNCDs, and personalized management to extend the healthy lifespan and reduce costs, is increasingly a global priority [1]. CNCDs include diseases that are not monogenic in nature, not purely environmental (trauma), and not purely somatic. They do include the most common forms of disease such as heart disease, metabolic disorders, neurological and mental health disorders, heritable cancers, and many non-congenital/non-monogenic pediatric disorders. Examples include myocardial infarction, arrhythmia, diabetes, Alzheimer's disease, prostate cancer, and autism spectrum disorder.

Recent advances in genotyping technology have greatly improved our understanding of the genetic risk factors that contribute to such diseases. In particular, whole-genome association studies have uncovered many common variants that increase an individual's risk of developing a disease during his/her lifetime. Since disease prevention will be the most effective means to ensure a healthier population in the coming decades, it is necessary to understand how to integrate inherited genetic risk information into our clinical decision-making process early in life so that we can minimize the chance of developing disease in the future. Low effect size common SNP variants, rare and private variants, DNA copy number variants and epigenetic modifications are thought believed to account for most of the inherited risk. When we can fully articulate the relative contribution of each of these elements to any specific disease, and the effects of their interactions with one another, our predictive accuracy will peak.

Accurately estimating an individual's risk to develop a CNCD is a challenging task. To begin, the risk is determined by many factors including the genetic risk factor load, environmental factors, gender, age etc and not all contributing factors are known. It is therefore clear that for most conditions the best risk assessments can only provide a probabilistic estimate. In order to accurately estimate the risk of an individual, one has to take into account the different associated variants, their effect sizes, their frequency in the population, the environmental factors affecting the individual, such as diet, age, family history and ethnic background as well as their interactions. Large-scale studies that investigate all of these factors at once are prohibitively expensive to conduct, and to our knowledge, none have been conducted.

Here, we study the performance of risk estimates based on the genetic composition of an individual alone, keeping all other factors fixed. Several approaches for risk estimation based on genetics alone have been proposed in the past [2–4]. These methods generally use the assumption that the disease-associated loci are independent of one another and that the relative risk of each locus is given. In practice, the relative risks are normally not known since in case-control studies, the odds-ratios and not the relative risks are given. In [2], the relative risks are inferred from the odds ratios by solving a set of equations that takes into account the prevalence of the disease, the frequencies of the genotypes and the odds ratios. Here, we suggest using a new method which aims at estimating the risk over the lifetime of an individual. The probability of disease as calculated using our method will be referred to as the Genetic Composite Index (GCI) or the GCI score (see Methods).

Similarly to previous approaches, we rely on several assumptions, main among them being the assumption of independence between the disease-associated loci. We use simulated data as well as real data to assess the performance of the risk estimates under different conditions. Importantly, we find that the assumption of independence does not greatly affect the generality of our method and modest SNP-SNP interactions in simulated data do not seem to significantly affect its predictability.

In order to measure the quality and effectiveness of GCI and similar methods, it is important to understand their limitations and merits. For example, [2] use Receiver Operating Characteristic (ROC) curves in order to measure the effectiveness of various risk measures. We adapt their use of ROC curves to evaluate our proposed score, and in particular, we consider the use of GCI in the context of three different diseases: Type 2 Diabetes, Crohn's disease and Rheumatoid Arthritis. We use simulations to calculate the predictive power of these different methods under an ideal "best-case" theoretical scenario, in which all the genetic factors are known. This ideal risk assessment depends on several factors including the heritability and the average lifetime risk of the disease. We find that the predictive power currently achieved for these diseases is substantially lower than the ideal predictive power, suggesting that major interactions and possible epigenetic factors are yet to be discovered. We emphasize that GCI is not a substitute for large-scale studies designed to simultaneously test multiple risk factors, but is rather an index that can be used when the result of such studies are simply not available, as is the case for virtually all common diseases.

## Results

### Evaluation of the GCI risk score and its assumptions

We use the Wellcome Trust Case Control Consortium (WTCCC) data [5] to test our GCI methodology. This dataset contains the genotypes of approximately 14,000 individuals divided into seven subpopulations based on disease phenotypes and one unaffected control subpopulation of 1,500 samples from the UK Blood Service Control Group. We limited our attention to the Type 2 Diabetes, Crohn's Disease and Rheumatoid Arthritis subpopulations and the common control group and did not consider any environmental variables in this analysis. We used SNPs that were reported to be significantly associated with each of these conditions in literature (see **Table 1**) and that passed a set of quality criteria. The main criteria were that i) The SNP association was consistently replicated within a given ancestral group and ii) The number of cases and controls were at least 250 when the effect size was less than 1.5 (Details about genotyping quality criteria in WTCCC data are given in [5]. There are no other criteria with respect to genotyping except that the SNPs chosen were reported in high quality studies that use good genotyping methodology). For each of the chosen SNPs, we computed the relative risk (see Methods) based on the empirical distribution of alleles found in the WTCCC dataset and used the GCI formula to calculate an estimated risk per individual. We note that some of the known risk variants are not present on the Affymetrix 500k GeneChip array that was used by the WTCCC, and therefore we expect the predictability of the GCI to be better than what is presented in our analysis below.

As noted before, we use Receiver Operating Characteristic (ROC) curve analysis [12] in order to evaluate the ability of GCI to serve as a predictive test for a condition. ROC curves have been previously used as a measure of the reliability of a genetics-based risk assessment test [2]. For a perfect test, a threshold $t$ could be chosen such that all individuals with a score larger than $t$ develop the condition, and all individuals with a score less than $t$ don't. However, in practice, we will find that for any given threshold there is some fraction of false positive and false negative assignments. The ROC curve graphically depicts the relationship between false positive rates and true positive rates, and thus it can be used to guide the tradeoffs between test sensitivity and specificity. We use the area under the ROC curve (AUC) as a quantitative measure to compare different risk scores. In general, the larger the value of the AUC, the better the score used for the classification. If classification were done randomly, the AUC is expected to be 0.5 and for the perfect score the AUC is equal to 1.

### Comparisons with an interactions model

One of the assumptions made by the GCI framework is that the disease-associated SNPs are independent. This assumption is useful since the score can then be calculated just from summary data; furthermore, when interactions are modeled based on limited data, there is a risk of over-fitting. Nevertheless, in an attempt to quantify how much information might be lost by the independence assumption, we compared our method with a model that accounts for both SNP-SNP interactions and the marginal contribution of each SNP. Particularly, we used logistic regression to account for the interactions. If the SNPs are $s_1, s_2 \ldots s_n$, then the model assumes that the logit transformation of the binary outcome reflecting disease or non-disease status is $X = c + a_1 s_1 + a_2 s_2 + \ldots + a_n s_n + a_{12} s_{12} + \ldots + a_{n-1,n} s_{n-1,n}$, where $s_{ij}$ is the interaction between $s_i$ and $s_j$. We first trained the model using the WTCCC data and then generated a ROC curve based on its probability estimates. Since this model takes into account the pairwise interactions between SNPs, it should be at least as accurate as the GCI score, which does not consider them. Note that the logistic regression model is an optimistic upper bound on the GCI since it can easily over-fit the model to the data; therefore, we are being conservative in our estimation of the information lost under the independence assumption. **Figure 1** shows the ROC curves

**Table 1.** Allele frequencies and the relative risks of Type 2 Diabetes, Crohn's Disease and Rheumatoid Arthritis SNPs.

| Disease | dbSNP rs id | Relative risk[1] for RR | Relative Risk[1] for RN | Frequency[2] of RR | Frequency[2] of RN |
|---|---|---|---|---|---|
| Type 2 Diabetes | rs10012946 [6] | 1.1464 | 1.0239 | 0.5000 | 0.4667 |
| | rs10811661 [7] | 1.3008 | 1.1282 | 0.6667 | 0.2500 |
| | rs1801282 [7] | 1.4128 | 1.2417 | 0.8667 | 0.1167 |
| | rs4402960 [7] | 1.1602 | 1.1233 | 0.1167 | 0.3500 |
| | rs4506565 [5] | 1.6133 | 1.2738 | 0.0847 | 0.3729 |
| | rs5215 [7] | 1.1681 | 1.0935 | 0.1000 | 0.6167 |
| | rs8050136 [8] | 1.3609 | 1.1176 | 0.1167 | 0.6667 |
| | rs9494266 [9] | 1.4909 | 1.2296 | 0.0169 | 0.0847 |
| | rs10923931 | 1.1948 | 1.0947 | 0.0167 | 0.2000 |
| | rs4607103 | 1.1392 | 1.0681 | 0.6333 | 0.3500 |
| | rs7961581 | 1.1355 | 1.0664 | 0.0500 | 0.3667 |
| | rs864745 | 1.1530 | 1.0747 | 0.3158 | 0.4035 |
| | rs5015480 | 1.1456 | 1.0451 | 0.3167 | 0.4833 |
| Crohn's Disease | rs10883365 | 1.6154 | 1.1989 | 0.3000 | 0.4000 |
| | rs2066845 | 11.4381 | 3.0164 | 0.0000 | 0.0333 |
| | rs10489276 | 1.4130 | 1.1888 | 0.0333 | 0.3667 |
| | rs1894603 | 1.4608 | 1.2088 | 0.2542 | 0.4407 |
| | rs4871611 | 1.1654 | 1.0795 | 0.3667 | 0.5000 |
| | rs6679677 | 1.7116 | 1.3085 | 0.7167 | 0.2833 |
| | rs17234657 | 2.3052 | 1.5360 | 0.0667 | 0.2000 |
| | rs11175593 | 2.3532 | 1.5353 | 0.0000 | 0.0333 |
| | rs11584383 | 1.3899 | 1.1790 | 0.4333 | 0.4500 |
| | rs1456893 | 1.4371 | 1.1989 | 0.3667 | 0.5333 |
| | rs1736135 | 1.3898 | 1.1790 | 0.3000 | 0.5000 |
| | rs17582416 | 1.3432 | 1.1590 | 0.1667 | 0.4333 |
| | rs2872507 | 1.2527 | 1.1193 | 0.2167 | 0.5000 |
| | rs3764147 | 1.5580 | 1.2484 | 0.0847 | 0.3220 |
| | rs4263839 | 1.4852 | 1.2188 | 0.4167 | 0.4667 |
| | rs744166 | 1.3898 | 1.1790 | 0.3276 | 0.4483 |
| | rs762421 | 1.2751 | 1.1292 | 0.2500 | 0.4833 |
| | rs10210302 | 1.8433 | 1.1890 | 0.3000 | 0.5000 |
| | rs7746082 | 1.3663 | 1.1690 | 0.1017 | 0.4915 |
| | rs7927894 | 1.3432 | 1.1591 | 0.2333 | 0.3833 |
| | rs9858542 | 1.8316 | 1.0895 | 0.0333 | 0.4167 |
| | rs11805303 | 1.8525 | 1.3875 | 0.1000 | 0.3833 |
| | rs1000113 | 1.9102 | 1.5354 | 0.0000 | 0.0667 |
| | rs2066844 | 3.2543 | 1.9609 | 0.0000 | 0.2203 |
| | rs17221417 | 1.9118 | 1.2883 | 0.1000 | 0.5167 |
| | rs2542151 | 1.9997 | 1.2980 | 0.0500 | 0.2833 |
| | rs10761659 | 1.5461 | 1.2287 | 0.2333 | 0.6333 |
| Rheumatoid | rs10118357 [10] | 1.7278 | 1.3152 | 0.2712 | 0.5254 |
| Arthritis | rs13207033 [10] | 1.7559 | 1.3258 | 0.6667 | 0.3167 |
| | rs6457617 [5] | 5.0847 | 2.3414 | 0.2167 | 0.5667 |
| | rs6679677 [11] | 3.1672 | 1.6847 | 0.0000 | 0.2833 |
| | rs6920220 [5] | 1.7023 | 1.1965 | 0.0000 | 0.3500 |

1. The relative risks provided here were calculated using the GCI methodology, as explained in the Methods section. RR means risk-risk genotype and RN means risk-nonrisk genotype.
2. The allele frequencies are taken from the HapMap project's CEU population.
doi:10.1371/journal.pone.0014338.t001

**A**

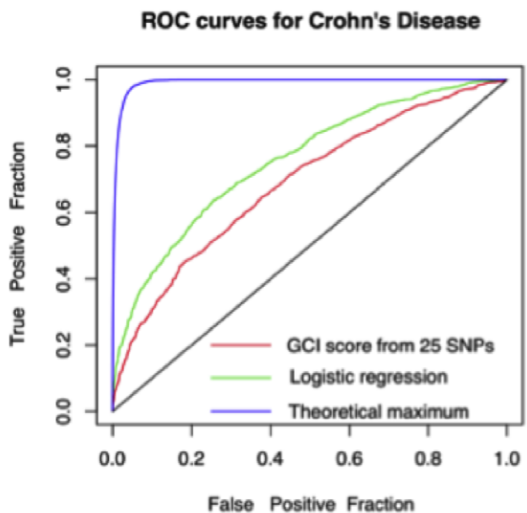

**B**

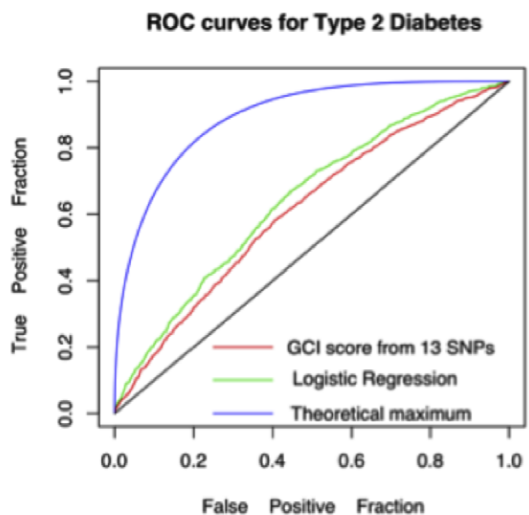

**C**

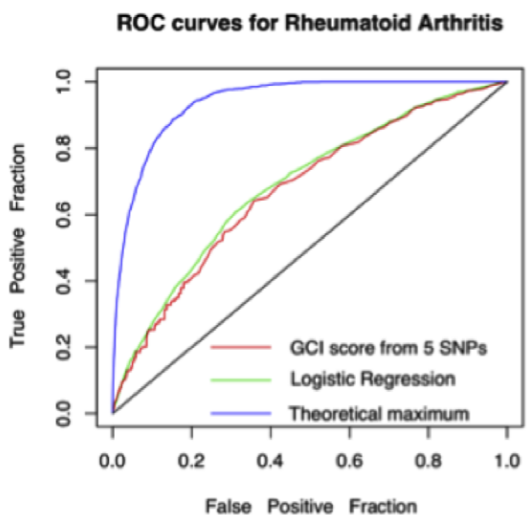

**Figure 1. ROC curves for the WTCCC dataset. A.** Crohn's Disease. **B**. Type 2 Diabetes. **C**. Rheumatoid Arthritis. In each plot, the black line corresponds to random expectation, the blue lines correspond to theoretical expectations (under the two disease models described in Methods) when the genetic variable is known, the red line corresponds to GCI, and the green line corresponds to logistic regression.
doi:10.1371/journal.pone.0014338.g001

for the three disease scenarios and **Table 2** gives their AUCs. We observe that the AUCs for GCI and logistic regression are quite similar for these three diseases, leading us to the conclusion that SNP-SNP interactions do not add substantial information to the risk assessment for the diseases investigated here. We can therefore justify our assumption (at least in these cases) that SNP-SNP interactions can be ignored as long as there is no evidence for such an interaction from previous studies.

## Theoretical upper bound for disease-risk predictability

The number of SNPs used in our analysis reflects the current knowledge about the effect of common SNPs on the risk of a disease. These, however lack many other factors such as epigenetic factors, rare variants, copy number variants, interactions etc. The question remains as to how much more accurate could we potentially be when considering genetic factors alone. We shed light on this by comparing our empirical results to theoretical disease models that assume that the disease is affected by both environmental and genetic factors, and that the two factors are independent (see Methods). Our model assumes that there are many small genetic effects that are cumulative and therefore the genetic factors include a normally distributed random variable. It takes into account the heritability and lifetime risk of the condition, resulting in a realistic extrapolation of the unknown genetic risk factors based on the currently known ones.

Formally, the theoretical model uses a phenotype variable P, and it assumes that $P = G + E$, where G is the genetic risk and E is the environmental risk and an individual will develop the condition in his/her lifetime if $P > \beta$ for a fixed $\beta$ (see Methods for more details). We generated 100,000 random samples for the distribution of P based on our theoretical models for G and E and determined their disease status. We then assumed that G is known for each individual (but E is unknown), and generated a ROC curve for the samples using this information alone. This curve represents an optimal scenario where the genetic risk is entirely understood and can be measured correctly for every individual but environmental risk factors are completely unknown. We will refer to the area under the ROC curve in this case as the theoretical genetic maximum. **Figure 1** shows the ROC curves for such a scenario and **Table 2** gives their areas. We observe that the GCI area under the curve with currently known variants is much less than that of the optimal theoretical genetic models, which suggests that many additional unknown genetic variants and/or interactions are expected to affect these diseases.

Based on **Figure 1**, we conclude that there is room for improvement in predictive modeling that will most likely come through the discovery of additional genetic variants and gene-environment interactions for the three conditions discussed in this text. It is useful to know what percentage of the genetic factors have been captured to date. Under the assumption that all the major genetic factors have already been discovered and that there are no gene-gene or gene-environment interactions, we can estimate the number of variants that will suffice to obtain a ROC curve with an AUC as large as the theoretical genetic maximum. If we assume that the GWAS studies performed to date have sufficient density to identify all large effect size common variants in

**Table 2.** The area under the ROC curve for the three diseases under three different scenarios.

| Disease | Heritability | Average Lifetime Risk | Optimal Scenario[1] | GCI score | Logistic Regression |
|---------|--------------|----------------------|--------------------|-----------|--------------------|
| Type 2 Diabetes | 64% [13] | 25.0% [16] | 0.894 | 0.613 | 0.644 |
| Crohn's Disease | 80% [14] | 0.56% [17] | 0.992 | 0.689 | 0.757 |
| Rheumatoid Arthritis | 53% [15] | 1.54% [18] | 0.944 | 0.675 | 0.689 |

1. The ideal score when the complete genetic information is known.
doi:10.1371/journal.pone.0014338.t002

the genome, and that all the unknown variants are common (minor allele frequency $= 10\%$), yet of weak effect size and that such variants contribute relative risks of 1.1 for the homozygous risk genotype and 1.05 for the heterozygous genotype; then our results show that under these assumptions the number of undiscovered risk factors is quite large (in the 1000s). Furthermore, we observe that only about 6% of the genetic variance is explained by the known variants for Type 2 Diabetes, about 9% for Crohn's disease, and about 14% for rheumatoid arthritis. It is also reasonable to assume that additional large-effect size variants will be discovered through the use of next-generation technologies and take the form of rare/*de novo* nucleotide variants, copy number variants and epigenetic modification of the primary nucleotide sequence – and that it is likely that a blend of a few of these larger effects will account for the missing heritability together with a larger number of common and weak effect size variants.

Attempts to estimate the number of causal variants in complex diseases have been made in the past [19–21]. These attempts reach somewhat different conclusions than ours, i.e. these studies estimate the number of genetic effects to be found to be quite modest, even under the assumption of independence between genes and environment. The main difference in the methodology between our approach and these previous approaches is that previous approaches have been published prior to the results achieved by GWAS studies. Thus, they do not make the assumption that the major common effects have already been found, and they do not take into account the heritability and lifetime risk. We note that [22] used a model similar to ours to investigate the relationship between the number of disease loci and the relative risk of the loci and their results are broadly similar to ours. They use the prevalence of the disease instead of lifetime risk. It must be mentioned that inaccuracies in the heritability estimates can affect these numbers, but as long as they are not off by an order of magnitude, we expect the results to be qualitatively similar.

## Theoretical effect of unknown SNP-SNP interactions

Our GCI score is based on the assumptions that all SNPs are in linkage equilibrium and that they have independent effects on the risk of the disease. As discussed above, the three examples studied here show no significant difference between the GCI model and a model in which pair-wise dependencies among the SNPs are included through logistic regression. This assumption may not always hold since, we know of some rare examples for which there is evidence of epistasis [23]. If these interactions are known, they can easily be incorporated into the GCI model by considering the interacting SNPs together as a combination. However, it is important to understand the effect of unknown SNP-SNP interactions on the multiplicative risk estimates.

In order to further explore the issue of interactions, we simulated datasets under a model in which a single pair of SNPs is interacting. Formally, the model can be described as follows. Let

$\lambda_i$ denote the relative risk of the disease for a particular combination of genotypes $(g_i)$ and p denote the average lifetime risk. If all SNPs are independent, the total risk is proportional to $\lambda_i = \prod_{j=1}^{n} \lambda_{ij}$ where $\lambda_{ij}$ denotes the relative risk for the $j^{th}$ locus. In the interactions model, we assume that for a particular pair, the relative risk for some combinations of genotypes is $\gamma$ times larger than the product of their relative risks. For all other SNPs and for all other genotype combinations, relative risks are assumed to be multiplicative. Thus, for example, if SNPs x and y interact, then the relative risk for the pair, $K = \gamma \lambda_{ix} \lambda_{iy}$ for certain configurations of $(g_{ix}, g_{iy})$, and $K = \lambda_{ix} \lambda_{iy}$ for other combinations. The total risk in this case would be $K \prod_{j \ne x, j \ne y}^{n} \lambda_{ij}$.

We set the values of $\lambda_{ix}$, $\lambda_{iy}$ for the interacting SNPs x and y so that the relative risks for each of these SNPs under univariate models is equal to what is observed in real data (given in **Table 1**). We assign the probability that an individual is a case to be $P(\text{disease} \mid g_i) = C\lambda_i$, where C is a normalizing factor, and $\lambda_i$ is the relative risk of individual i based on the interactions model. We choose C so that the fraction of cases is close to the average lifetime risk of the disease.

Let RR, RN and NN denote the observed values of relative risks for any SNP for risk-allele homozygote (2), heterozygotes (1) and non-risk-allele homozygote (0) respectively and let rr, rn and nn denote the respective genotype frequencies. Since $\lambda_{ij}$ for any locus j can only take 3 possible values corresponding to the 3 possible genotypes, we will denote these by $\lambda_{ij0}$, $\lambda_{ij1}$, and $\lambda_{ij2}$ respectively and set $\lambda_{ij0} = 1$ for all SNPs. We obtain values of $\lambda_{ix1}$, $\lambda_{iy1}$, $\lambda_{ix2}$, $\lambda_{iy2}$ for SNPs x and y by solving the following system of equations:

$$RR_x = (\gamma rr_y \lambda_{ix2} \lambda_{iy2} + \gamma rn_y \lambda_{ix2} \lambda_{iy2} + nn_y \lambda_{ix2})/(rr_y \lambda_{iy2} + rn_y \lambda_{iy1} + nn_y)$$

$$RN_x = (\gamma rr_y \lambda_{ix1} \lambda_{iy2} + \gamma rn_y \lambda_{ix1} \lambda_{iy1} + nn_y \lambda_{ix1})/(rr_y \lambda_{iy2} + rn_y \lambda_{iy1} + nn_y)$$

$$RR_y = (\gamma rr_x \lambda_{iy2} \lambda_{ix2} + \gamma rn_x \lambda_{iy2} \lambda_{ix1} + nn_x \lambda_{iy2})/(rr_x \lambda_{ix2} + rn_x \lambda_{ix1} + nn_x)$$

$$RN_y = (\gamma rr_x \lambda_{iy1} \lambda_{ix2} + \gamma rn_x \lambda_{iy1} \lambda_{ix1} + nn_x \lambda_{iy1})/(rr_x \lambda_{ix2} + rn_x \lambda_{ix1} + nn_x)$$

Based on the risks in the interactions model, we assigned disease status labels for 100,000 randomly drawn samples. We used this

**A**

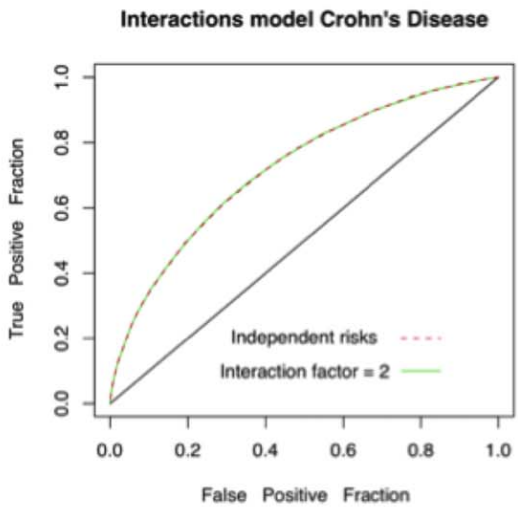

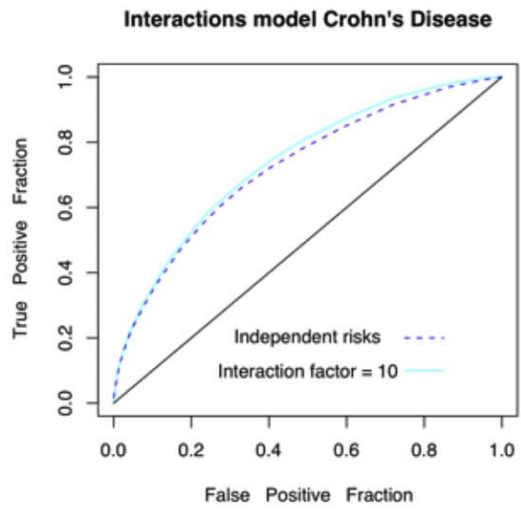

**B**

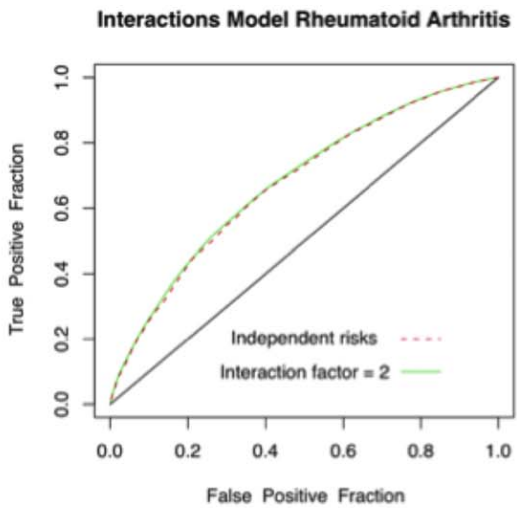

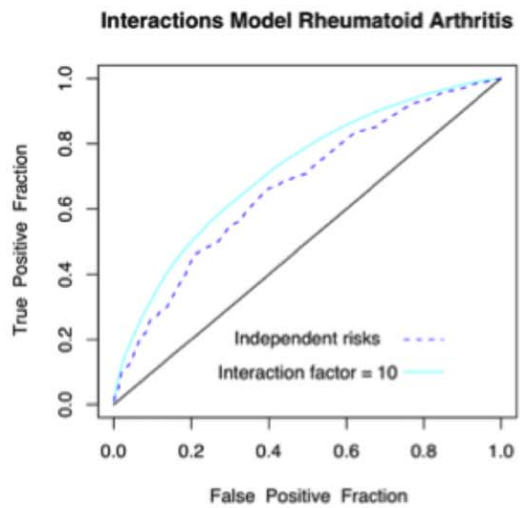

**C**

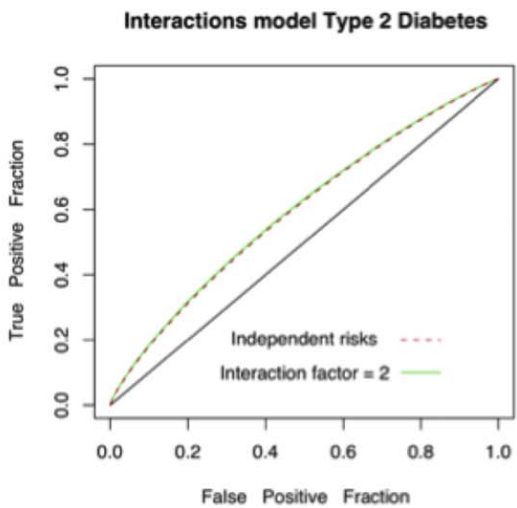

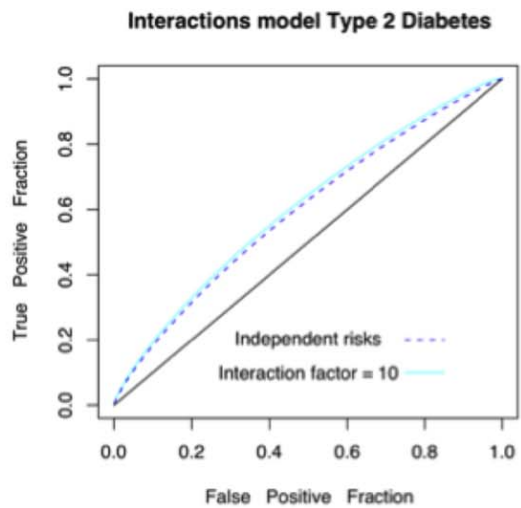

**Figure 2. ROC curves for models with interactions vs the simple multiplicative model. A**. Crohn's Disease. **B**. Rheumatoid Arthritis. **C**. Type 2 Diabetes. In each plot, 1,000 threshold points were used.
doi:10.1371/journal.pone.0014338.g002

simulated case-control data to plot ROC curves based on two approaches for risk assessment. First, we calculate the relative risk of an individual according to the true interactions model. Then, we assigned relative risks assuming the independence model. As observed in **Figure 2** and in **Table 3**, we find that the ROC curves can differ marginally when the interaction factor is high (i.e. $\gamma = 10$). However, it can be argued that strong deviations from the independence model will also be more detectable in genome wide association studies. Particularly, whole-genome association studies often report that SNP-SNP interactions were tested but were not found to be significant (e.g. [24]). Therefore, when no interactions have been reported in the literature for a set of SNPs, it seems unlikely that the classification accuracy of the multiplicative test will differ dramatically from that of the true model that includes interactions.

### Measuring the Absolute Error in the Risk Estimate

The ROC curve serves as one metric for evaluating a diagnostic in that it provides a quantitative measure of the ability of the test to distinguish between unaffected and affected individuals. However, when estimating the lifetime risk, the ROC curve alone may not be sufficient if a score does not directly estimate the correct probabilistic measure (i.e. the probability of developing disease in one's lifetime) but instead computes some function of this probability. In particular, for any given pair of score functions, $f_1(G)$ and $f_2(G)$, the ROC curves of the functions will be identical as long as $f_1$ is a monotonic increasing function of $f_2$. For instance, we could simply assign $f_2(G) = \log(f_1(G))$, and in this case by using the scores $f_1$ and $f_2$ to estimate risk we will get exactly the same ROC curves. However, these two functions may give very different lifetime risk estimates to individuals. Therefore, ROC curves alone are not sufficient for tests that report probabilistic risk. For quality assessment, we also need a more informative quantity, the absolute value of relative error between the true risk probability and the estimated risk probability. The relative error is defined as the difference between the estimated and true risk probability divided by the true risk probability. Thus, the absolute value of relative error is given by:

$$|\text{Estimated Risk Probability} - \text{True Risk Probability}|/$$
$$\text{True Risk Probability}$$

Since the true probability of developing a disease is unknown, we simulated a scenario in which case-control data is used to calculate the GCI parameters (i.e. the relative risks), and then applied the GCI risk estimates to another independently simulated population. The disease model we used for the simulation assumes that the genetic factors of the disease can be decomposed into a small number of large effects and a large number of small effects that can be approximated by a normal distribution (see Methods). Since most diseases are diagnosed later in life, we introduced the age of onset of the disease to the model. For each individual that has been determined to develop the disease based on the model, we choose the age of onset of the disease based on some distribution for the age of onset (Normal distribution with mean $= 50$ and SD $= 13$). Thus, in our simulation, some of the controls may in fact be cases that have not been diagnosed at the time of the study. To create a realistic simulation of an age-matched case-control study, we first repeatedly simulated the genetic and environmental factors, as well as the age of onset for individuals; we picked the age of the individuals from a uniform distribution between 0 and 100. We generated 10,000 cases using this process. For each of these cases, we generated an age-matched control by sampling 10,000 controls conditioned on their age. We estimated the odds ratios for each SNP based on this case-control data, and then used these odds ratios to calculate the relative risks for each SNP associated with the disease, using our GCI methodology.

The above procedure was used to generate a simulated set of relative risk values. We then generated 500 individuals randomly according to the theoretical disease model. Since the variables are known for each of these individuals, we know the correct genetic risk to develop the condition. We use these 'true risks' as a baseline for the accuracy measure. We compare the GCI based risk estimates to this baseline, as well as a variant of the GCI in which the relative risks are replaced by the odds ratios. We note that methods that calculate disease risk based on prevalence (e.g. [2]) will usually get relative risks that are close to the odds ratios.

In **Figure 3**, we plot the distribution of the absolute value of relative errors for a simulated disease with average lifetime risk of 25% and heritability of 64% (**Figure 3a**), and for a disease with average lifetime risk of 42% and heritability of 57% (**Figure 3b**). These values roughly correspond to the lifetime risk and heritability of Type 2 Diabetes and Myocardial Infarction respectively. It is clear from the Figure that there is a dramatic difference between the lifetime risks when using the relative risks and when using the odds ratios. This may not be noticeable using a ROC curve that only measures the classification accuracy. Thus, using odds ratios or prevalence based calculation for risk relative

**Table 3.** The area under the curve (AUC) for the different interaction scenarios.

| | Simulated Interaction Factor 2[1] | | Simulated Interaction Factor 10[2] | |
|---|---|---|---|---|
| | Interaction risk estimate | GCI risk estimate (Multiplicative) | Interaction risk estimate | GCI risk estimate (Multiplicative) |
| Crohn's Disease | 0.722 | 0.722 | 0.739 | 0.724 |
| Rheumatoid Arthritis | 0.679 | 0.674 | 0.720 | 0.673 |
| Type 2 Diabetes | 0.597 | 0.594 | 0.607 | 0.595 |

1. The two columns correspond to the case where there is a SNP-SNP interaction in which the effect of a certain combination of genotypes has two times the product of the marginal effects.
2. The two columns correspond to the case where there is a SNP-SNP interaction in which the effect of a certain combination of genotypes is 10 times the product of the marginal effects.
doi:10.1371/journal.pone.0014338.t003

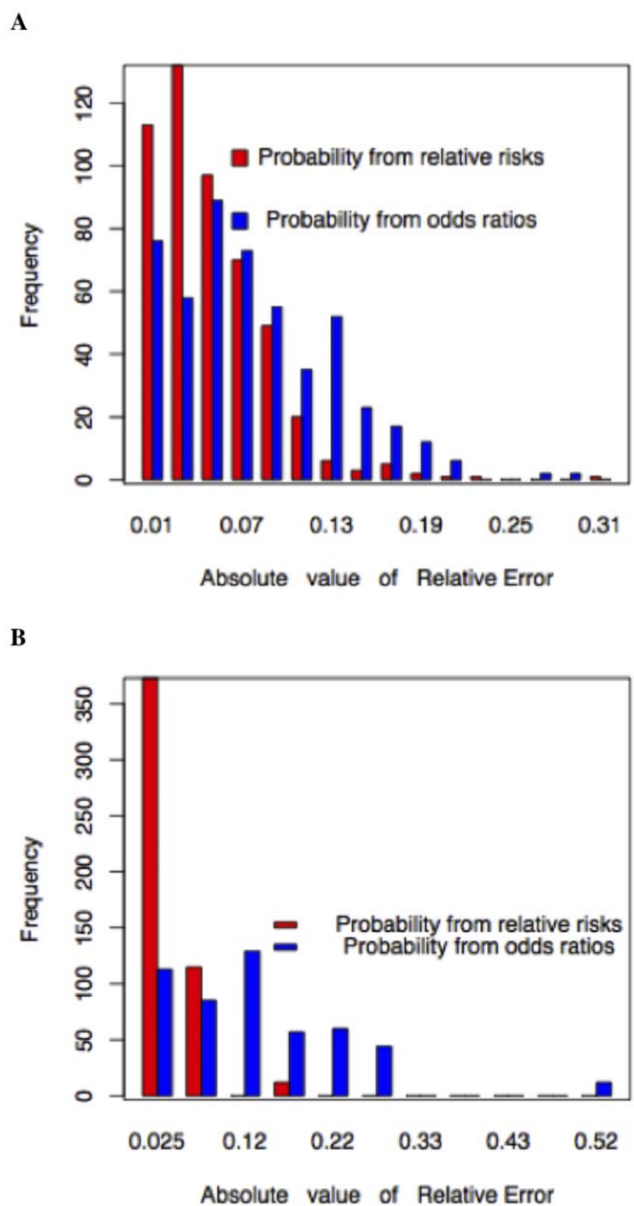

**Figure 3. Relative errors for the estimated lifetime risk probabilities. A.** Comparison of odds ratios and relative risks for Type 2 Diabetes with lifetime risk of 25% and heritability of 64%. **B.** Comparison of odds ratios and relative risks for Myocardial Infraction with lifetime risk of 42% and heritability of 57%.
doi:10.1371/journal.pone.0014338.g003

generally inflates the results for lifetime risk calculations, and under some circumstances can generate lifetime risk estimates that are larger than 100% (hence these are not good enough for lifetime risk calculations and our methodology is necessary).

## Can the addition of environmental risk factors improve our current predictions?

In the previous sections, we used only the genetic information to estimate the risk of disease. In order to estimate the potential contribution of known environmental factors to disease prediction, we now consider the case where both environmental and genotypic data are used to estimate risk. Such an example was studied for the case of Type 2 Diabetes in [2]. Here, we demonstrate the utility of environmental factors across Type 2 Diabetes, Crohn's Disease and Rheumatoid Arthritis, which have very different heritability and average lifetime risk values. As in [2], we also assume that the risks and frequencies across all SNPs as well as across all environmental factors are independent and multiplicative. Based on this assumption, we generalized the GCI score for the case where environmental factors are also taken into account. We call the resulting statistic for lifetime risk Environmental-Genetic Composite Index (EGCI). The EGCI score (like the GCI score) is defined as the product of relative risks across all the markers and all of the environmental factors normalized by a constant. Note that when calculating the EGCI, the different levels/classes of any particular environmental factor are treated in exactly the same way as the different alleles of a marker in GCI. Thus, environmental factors are mathematically no different from additional markers. **Table 4** gives the frequencies and relative risks of the environmental variables for the 3 diseases.

We simulated the genotype and environmental factor values for a set of 100,000 individuals based on their known frequencies in the population (See **Tables 1 and 4**). For every individual, we randomly and independently generate each genotype and environmental variable using these frequencies (In particular, we use a uniform random number between 0 and 1 for doing this). We then randomly assigned a disease status for all individuals based on the lifetime risk probabilities calculated from the generalized multiplicative model (i.e. EGCI). Next, we compared the predictive power of the pure genetics based GCI score to the new generalized EGCI score. The ROC curves for Type 2 Diabetes, Crohn's Disease and Rheumatoid Arthritis are shown in **Figure 4**. The added value of environmental factors is not dramatic for Crohn's Disease and Rheumatoid Arthritis, however it is substantial for Type 2 Diabetes. This is driven by the fact that Body Mass Index is crucially affecting the risk for Type 2 Diabetes (with a relative risk of 42.1 when BMI >35 [25]). Note that for a disease such as Crohn's disease we do not expect environmental factors to play a major role since the heritability of this condition is roughly 80%.

## GCI and EGCI for Type 2 Diabetes case-control data from the GENEVA study

GENEVA study refers to the Gene Environment Association Studies initiative (www.genevastudy.org) funded by the trans-NIH Genes, Environment, and Health Initiative (GEI). The goal of this study is to identify novel genetic factors that contribute to Type 2 Diabetes Mellitus through a large-scale genome-wide association study of well-characterized cohorts of nurses and health professionals. In this study, around 1 million SNPs have been genotyped in about 2712 cases with Type 2 Diabetes and 3179 controls. A variety of environmental variables have also been collected for these individuals. We illustrate the performance of GCI and EGCI methodology using 15 disease SNPs present in the GENEVA dataset. We only used unrelated individuals of Caucasian ancestry for this analysis. For calculating EGCI, we considered 2 environmental variables namely the Body Mass Index (BMI) and the smoking status (**Table 5** gives their relative risks). The results obtained are shown in **Figure 5** and the SNPs used are listed in **Table 6**.

## Discussion

The Human Genome Project [26], the HapMap project [27], and related initiatives have resulted in a reference human genome sequence, a catalog of common genetic variation and a haplotype map of several reference populations. Furthermore, this

**Table 4.** Relative risks of environmental variables for Type 2 Diabetes, Crohn's disease and Rheumatoid Arthritis.

| Disease | Environmental Variable | Level | Proportion in the population | Relative risk |
|---|---|---|---|---|
| | | <23 | 0.20 | 1.00 |
| | | 23–23.9 | 0.16 | 1.00 |
| | | 24–24.9 | 0.14 | 1.50 |
| | | 25–26.9 | 0.27 | 2.20 |
| Type 2 Diabetes | Body Mass Index | 27–28.9 | 0.14 | 4.40 |
| | | 29–30.9 | 0.06 | 6.70 |
| | | 31–32.9 | 0.02 | 11.6 |
| | | 33–34.9 | 0.01 | 21.3 |
| | | >=35 | 0.01 | 42.1 |
| | | Never Smoked | 0.50 | 1.00 |
| | Smoking | Ex-Smoker | 0.39 | 1.10 |
| | | <20 cigs/day | 0.04 | 1.50 |
| | | >=20 cigs/day | 0.07 | 1.70 |
| | | Never Smoked | 0.545 | 1.00 |
| Crohn's Disease | Smoking | Ex-Smoker | 0.245 | 1.70 |
| | | Current-Smoker | 0.198 | 3.00 |
| | | Never Smoked | 0.498 | 1.00 |
| Rheumatoid Arthritis | Smoking | Ex-Smoker | 0.276 | 1.40 |
| | | Current-Smoker | 0.227 | 1.30 |

doi:10.1371/journal.pone.0014338.t004

information combined with cost-effective technologies to test associations between variations throughout the genome and traits and diseases of all sorts, has resulted in dozens of common variants shown to be unequivocally statistically associated with the risk of common diseases. These common variants can be used much like population-derived environmental risk factor data in assessing probabilistic pre-symptomatic risk of disease.

We have presented a new method for the estimation of an individual's lifetime risk based on genetic data through a genetic score function (the GCI). The GCI, like all estimates of a particular quantity, requires a set of assumptions that may bias the risk estimates. Particularly, the assumptions made by the GCI score are that the allele frequencies of the causal SNPs and effect sizes are known, and that all the SNPs are independent of each other. We show through simulation studies and by the analysis of the WTCCC data that, moderate SNP-SNP interactions have almost no effect on the power of the multiplicative GCI score. However, in principle strong non-additive effects between variants might affect the risk estimates, and thus care has to be taken when interpreting the results. In most scenarios, we expect that such effects will likely be discovered prior to the use of GCI and can be incorporated in the risk calculation. So, we view this as a minor problem, especially given that no significantly strong SNP-SNP interactions have been uncovered in whole genome association studies performed over the past several years.

We used the ROC curve analysis and the heritability of each of the conditions we considered to find the total genetic variation explained by known variants, compared to the expected genetic variation based on heritability. We find that current scientific knowledge can explain approximately 6%–14% of the total genetic variation for these conditions. This suggests that the risk estimates provided by the GCI may vary considerably in the future, as more genetic variants are found and used for risk estimation (e.g. see

[24]). The fact that only a small fraction of the genetic variants have been found to date suggests that the variance of the risk calculated by the GCI is still large; however, the GCI score aims at estimating the expected frequency of individuals with a given genetic load that will develop the condition during their lifetime, and the accuracy of the estimate of expectation will not be affected by the number of unknown variants.

It is clear that next-generation technologies will be used in study designs similar to GWAS to identify additional heritable risk factors for CNCDs. As each new genetic association is validated to the appropriate industry thresholds, this new genetic risk information can be added into the GCI in a scalable fashion, on a disease-by-disease basis to improve the accuracy of the GCI in real time.

Given these interpretations of the GCI score, it is informative to use such a score in order to estimate the risk of an individual based on their genetic data. The medical benefits of such individualized knowledge are intuitive, but have to be clinically proven through prospective studies. The main open question is whether individuals will benefit by change of behavior, early diagnosis or an individualized course of treatment based on their genetic information for actionable CNCDs. We believe that tools such as the GCI score will facilitate such studies and help transition us into the era of personalized preventive medicine.

## Methods

### Ethics Statement

The datasets used were approved by the relevant boards in Navigenics Inc and University of California Davis.

### Introduction

We consider a disease for which k risk loci have been identified. As done in [2,3], we assume that the different loci are acting

**A**

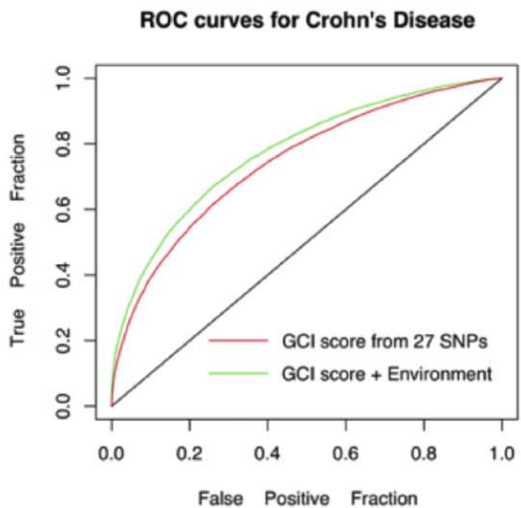

**B**

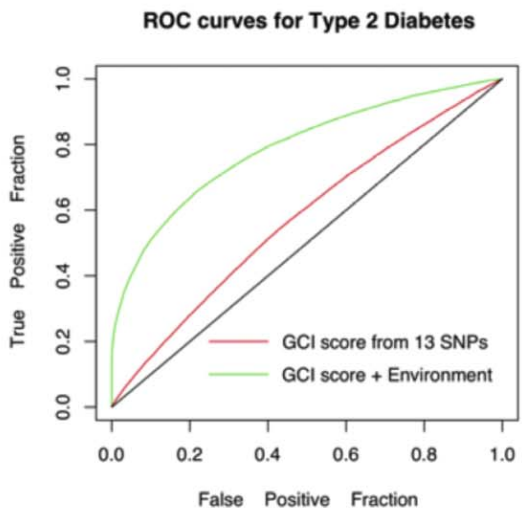

**C**

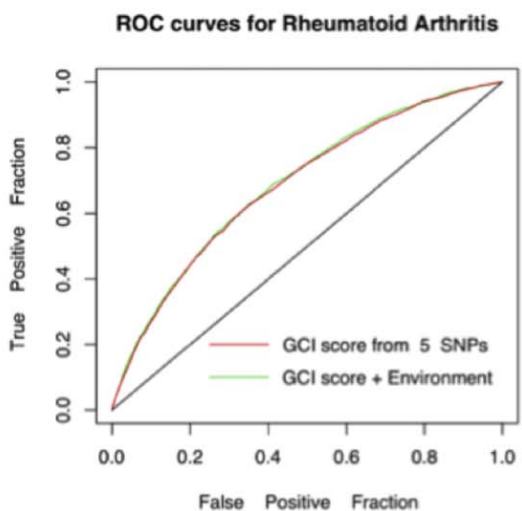

**Figure 4. EGCI vs. GCI in simulated data. A.** Effect of known genetic and environmental factors versus known genetic factors alone for Crohn's Disease. The AUCs of the two curves are 0.74 and 0.78. We considered smoking as the environmental variable in addition to the genetic factors. **B.** Effect of known genetic and environmental factors versus known genetic factors alone for Type 2 Diabetes. The AUCs of the two curves are 0.58 and 0.79 respectively. We considered Body Mass Index, alcohol intake and smoking frequency as the environmental factors for Type 2 Diabetes, in addition to the genetic factors. **C.** Effect of genetic and environmental factors versus genetic factors alone for Rheumatoid Arthritis. The AUCs of the two curves are 0.685 and 0.690. We considered smoking as the environmental variable in addition to the genetic factors. The relative risks for the environmental variables are provided in Table 4.
doi:10.1371/journal.pone.0014338.g004

independently, and thus $\Pr(g_1,\ldots,g_k|D) = \prod_{i=1}^{k} \Pr(g_i|D)$, and $\Pr(g_1,\ldots,g_k) = \prod_{i=1}^{k} \Pr(g_i)$, where $g_i$ is genotype of an individual in locus i, and $D$ represents the event that the individual will develop the disease across his or her lifetime. As noted by [2], it is straightforward to extend this model to cases where some interactions are known. Previous methods consider $D$ as the event that the individual is currently diseased and thus the risk estimated by these methods is for a snapshot in time. Such risk is related to the overall lifetime risk of the disease but with obvious differences. This difference can be quite dramatic in some cases, as we show in the results section.

When calculating the risk across multiple SNPs for an individual with genotypes $(g_1,\ldots,g_n)$, we are interested in finding the probability $\Pr(D|g_1,\ldots,g_n)$. Using Bayes law and the independence assumption

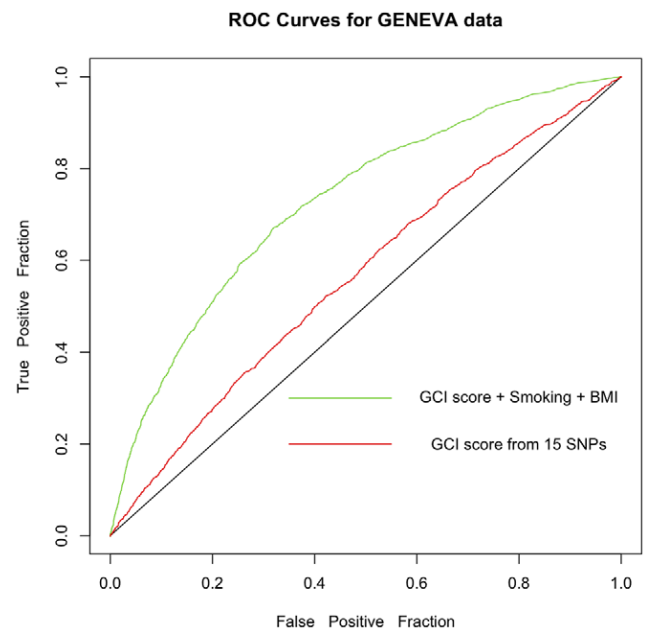

**Figure 5. ROC curves for the GENEVA dataset.** Effect of genetic (15 SNPs given in Table 6) and environmental factors (BMI, Smoking) versus genetic factors alone for predicting Type 2 Diabetes in 2600 cases and 3000 controls in the GENEVA data. The AUCs of the two curves are 0.727 and 0.565 respectively. The relative risks for BMI and Smoking are given in Table 5.
doi:10.1371/journal.pone.0014338.g005

**Table 5.** Relative risks of environmental variables for Type 2 Diabetes.

| Disease | Environmental Variable | Level | Relative risk |
|---|---|---|---|
| | | <23 | 1.00 |
| | | >=23 and <25 | 2.67 |
| | Body Mass Index | >=25 and <30 | 7.59 |
| | | >=30 and <35 | 20.1 |
| Type 2 Diabetes | | >=35 | 38.8 |
| | | Never Smoked | 1.00 |
| | Smoking | Ex Smoker | 1.23 |
| | | Current Smoker | 1.44 |

doi:10.1371/journal.pone.0014338.t005

$$\Pr(D|g_1,\ldots,g_n) = \frac{\Pr(g_1,\ldots,g_n|D)\Pr(D)}{\Pr(g_1,\ldots,g_n)}$$
$$= \frac{\Pr(D)\prod_{i=1}^{n}\Pr(g_i|D)}{\prod_{i=1}^{n}\Pr(g_i)} = \frac{\prod_{i=1}^{n}\Pr(D|g_i)}{\Pr(D)^{n-1}}$$

In order to estimate the lifetime risk of a specific individual, we therefore need to have an estimate of the average lifetime risk $\Pr(D)$ across the entire population and the risk of developing the disease across the lifetime of an individual with genotype $g_i$. The former has been estimated for a wide range of conditions using prospective studies [16–18]. The latter can be estimated using our method from case-control studies as described below.

## Odds ratios vs. relative risk

In epidemiology literature, the relative risk is often considered an intuitive and informative measure of risk. The relative risk is defined as $\lambda_i = \dfrac{\Pr(D|a_i)}{\Pr(D|a_0)}$, where $a_0$, $a_1$, and $a_2$ correspond to the genotypes with 0, 1, and 2 risk alleles. If the relative risks are known, we could estimate $\Pr(D|a_i)$ by using the following:

$$\Pr(D) = \Pr(D \mid a_2)\Pr(a_2) + \Pr(D \mid a_1)\Pr(a_1) + \Pr(D \mid a_0)\Pr(a_0) \quad (1)$$

Equation 1, together with the relative risks provide three independent equations with three variables, since $\Pr(a_i)$ can be found by considering a reference population, and $\Pr(D)$ is known. Unfortunately, the relative risk cannot be directly calculated in the context of case-control studies and whole-genome association studies. The relative risk can usually be estimated through prospective studies in which a set of healthy individuals is studied over a long period of time. In contrast, odds ratios are normally reported in case-control studies. The odds-ratio is the ratio between the odds of carrying the risk allele in cases vs. controls. For rare diseases, the odds ratio is a good approximation of relative risk; however for common diseases, the odds ratio could result in a misleading estimate of risk, where the odds ratios may be quite high even when the increase in risk is minor.

As previously noted [2], one can estimate the relative risks from the odds ratios by solving a set of equations. However, the equations proposed in [2] assume that the control population will never develop the disease. In the context of lifetime risk estimation this assumption is no longer valid since a subset of the control population might eventually develop the disease.

**Table 6.** SNPs used when analyzing the GENEVA genotype data.

| Disease | dbSNP rs id | Relative risk[1] for RR | Relative risk[1] for RN | Frequency[2] of RR | Frequency[2] of RN |
|---|---|---|---|---|---|
| Type 2 Diabetes | rs153143 | 1.1586 | 1.0772 | 0.0170 | 0.1670 |
| | rs11634397 | 1.0961 | 1.0472 | 0.3280 | 0.5340 |
| | rs8042680 | 1.1112 | 1.0545 | 0.0330 | 0.3670 |
| | rs10012946 [6] | 1.1464 | 1.0239 | 0.5000 | 0.4667 |
| | rs10811661 [7] | 1.3008 | 1.1282 | 0.6667 | 0.2500 |
| | rs1801282 [7] | 1.4128 | 1.2417 | 0.8667 | 0.1167 |
| | rs4402960 [7] | 1.1602 | 1.1233 | 0.1167 | 0.3500 |
| | rs4506565 [5] | 1.6133 | 1.2738 | 0.0847 | 0.3729 |
| | rs5215 [7] | 1.1681 | 1.0935 | 0.1000 | 0.6167 |
| | rs8050136 [8] | 1.3609 | 1.1176 | 0.1167 | 0.6667 |
| | rs10923931 | 1.1948 | 1.0947 | 0.0167 | 0.2000 |
| | rs4607103 | 1.1392 | 1.0681 | 0.6333 | 0.3500 |
| | rs7961581 | 1.1355 | 1.0664 | 0.0500 | 0.3667 |
| | rs864745 | 1.1530 | 1.0747 | 0.3158 | 0.4035 |
| | rs5015480 | 1.1456 | 1.0451 | 0.3167 | 0.4833 |

1. The relative risks provided here were calculated using the GCI methodology, as explained in the Methods section. RR means risk-risk genotype and RN means risk-nonrisk genotype.
2. The allele frequencies are taken from the HapMap project's CEU population.
doi:10.1371/journal.pone.0014338.t006

## Calculating Risk in the Presence of Diseased Controls

We now turn to the calculation of $\Pr(D|g_i)$ given that an $\alpha$ fraction of the controls will eventually develop the disease along their lifetime. We consider a locus in which $m+1$ different alleles are present. This allows us to deal with general scenarios, in which $g_i$ may represent any number of interacting SNPs, and where $m = 3^s$, where $s$ is the number of SNPs represented by $g_i$.

We will denote the $m+1$ possible alleles by $a_0, a_1\ldots, a_m$, where $a_0$ is the non-risk allele, and their respective allele frequencies in the general population as $f_0, f_1,\ldots, f_m$. Given that an $\alpha$ fraction of the controls will eventually develop the condition, we can write the odds ratios as:

$$OR_i = \frac{\Pr(D|a_i)(\alpha\Pr(D|a_0)+(1-\alpha)(1-\Pr(D|a_0))}{\Pr(D|a_0)(\alpha\Pr(D|a_i)+(1-\alpha)(1-\Pr(D|a_i))}$$

$$= \frac{\Pr(D|a_i)((2\alpha-1)\Pr(D|a_0)+1-\alpha)}{\Pr(D|a_0)((2\alpha-1)\Pr(D|a_i)+1-\alpha)}$$

From this, we get: $\Pr(D|a_i) =$

$$\frac{(1-\alpha)OR_i\Pr(D|a_0)}{1-\alpha+(1-2\alpha)\Pr(D|a_0)(OR_i-1)} \quad (2)$$

Similar to Equation 1, we know that $\Pr(D)=\sum_{i=0}^{m} f_i\Pr(D|a_i)$. Therefore, by Equation 2, we get the following:

$$\Pr(D) = \sum_{i=0}^{m} \frac{f_i(1-\alpha)OR_i\Pr(D|a_0)}{1-\alpha+(1-2\alpha)\Pr(D|a_0)(OR_i-1)}$$

For a fixed $\alpha$, we can solve this equation using a binary search on the variable $\Pr(D|a_0)$; there is exactly one solution between 0 and $\Pr(D)$ since the right hand side of this equation is an increasing function of $\Pr(D|a_0)$ and binary search is guaranteed to find that solution.

Generally, the value of $\alpha$ is unknown and it has to be determined based on the age characteristics of the study population. For instance, if the control population is a sample from the general population, then $\alpha$ should be taken as the average lifetime risk of the disease. However, if the control population was chosen so that their age range is after the age of onset of the disease, $\alpha$ should be close to 0. When case-control genotype data is given, one can use maximum likelihood estimation to calculate $\alpha$.

## Calculating the GCI score

The GCI method essentially provides a way to compute the relative risks of an individual as compared to an individual with non-risk alleles at each of the disease-associated marker. In order to calculate the lifetime risk, we take the product of the relative risks across all loci (this is the overall relative risk of the individual under the multiplicative model) and multiply it by the average lifetime risk of the disease in the population. We then divide this product by the average overall relative risk of the population. To approximate the average relative risk of the population, we assume that the SNPs at different loci are independent of one another (i.e. in linkage equilibrium). Under this assumption, the average overall relative risk of the population is equal to the product of the average relative risks at each disease-associated marker.

If all the markers effects are independent, the relative risk of individual $i$ is equal to $\lambda_i = \prod_{j=1}^{n} \lambda_{ij}$ where $\lambda_{ij}$ denotes the relative risk for the $j^{th}$ locus. Let $\Pr(D)$ denote the average lifetime risk of the disease in the population. Then, the GCI lifetime risk probability or GCI score of an individual $i$ is:

$$\Pr(D)\prod_{j=1}^{n} \lambda_{ij} / \prod_{j=1}^{n} \left( \sum_{k=0}^{k=m} f_{jk}\lambda_{jk} \right)$$

Here, $m+1$ alleles are possible at each marker locus and $\lambda_{jk}$ denotes the relative risk of the $k^{th}$ allele of the $j^{th}$ locus and $f_{jk}$ denotes its frequency in the sample.

## Theoretical Disease Models

We compared the GCI score to the optimal risk scores calculated under two different theoretical disease models. These models assume that the disease is affected by both environmental and genetic factors, and that the two factors are independent of each other. We denote the phenotype $P = G + E$, where $G$ is the genetic variable and $E$ is the environmental variable. Our first model assumes that both $G$ and $E$ are normally distributed with standard deviations of $\sigma_G$ and $\sigma_E$ respectively, and that an individual will develop the condition in his or her lifetime if $P > \alpha$ for a fixed $\alpha$. Similar models have often been used when heritability calculations are made [28]. We fix $\sigma_G$, $\sigma_E$ and $\alpha$ using the constraint that $h = \sigma_G^2/(\sigma_G^2 + \sigma_E^2)$, and that the average lifetime risk is equal to Probability $(P > \alpha)$. Since the heritabilities and average lifetime risks are known for each of the conditions we test, we can set the parameters of the models according to the disease. For this disease model, we can analytically show that the theoretical genetic maximum of AUC (i.e. when $G$ is known but $E$ is unknown) only depends on the heritability and the average lifetime risk of the disease (See next section) and not on the choice of $\sigma_G$, $\sigma_E$, or $\alpha$ which are difficult to estimate.

In the second model, a variant of the previous model, we assume that $G = \sum \lambda_i X_i + G1$, where $G1$ is normally distributed with standard deviation $\sigma_{G1}$, and $X_i \sim B(2, p_i)$ is Binomially distributed. In this case, $X_i$ corresponds to SNPs with large effects and $G1$ represents many other small genetic effects; if there are enough small genetic effects, we expect that the asymptotic behavior of their sum would be according to a normal distribution. By setting the parameters $\lambda$, $\sigma_{G1}$ and $p$ appropriately, we can control the relative risks of the large effect SNPs. We tune these parameters such that the relative risks are close to values observed in Table 1 (see below). As for the previous model, we can show that when $G$ is known (but $E$ is unknown) and the relative risks of the large effect SNPs and risk-allele frequencies are fixed, the area under the ROC curve for the second model only depends on the heritability and the average lifetime risk of the disease (see below).

## Proof for theoretical disease model 1

In this section, we will show that the theoretical genetic maximum of the area under the ROC curve for model 1 depends on the average lifetime risk (ALTR) and the heritability of the disease alone. Let $\sigma_e$ denote the variance in the environmental variable and $\sigma_g$ denote the variance in the genetic variable. In model 1, both genetic $(G)$ and environmental $(E)$ variables are normally distributed. The theoretical maximum of ROC curve is obtained when the genetic variable is known exactly while the environmental variable is unknown. An individual is a true case if $G + E > \alpha$ and a true control otherwise. For any cutoff chosen for the genetic variable, the individuals who are above that cutoff will be counted as cases and the rest as controls. The true positive fraction (TPF) is the fraction of true cases that are called as cases and false positive fraction (FPF) is the fraction of true controls that are called as cases. The TPF versus FPF for different values of cutoffs gives us the ROC curve.

The probability that an individual's genetic variable is greater than some cutoff (c) is given by: $P(G > c) = \int_{\beta\sigma_g}^{\infty} e^{-x^2/2\sigma_g^2} dx/\sqrt{2\pi}\sigma_g$ where $\beta = c/\sigma_g$.

The probability that an individual's genetic variable is greater than the cutoff and the individual is a true case is: $P(G > c$ and $G + E > \alpha) = \int_{\beta\sigma_g}^{\infty} e^{-x^2/2\sigma_g^2}(\int_{\gamma\sqrt{\sigma_g^2+\sigma_e^2}-x}^{\infty} e^{-y^2/2\sigma_e^2} dy/\sqrt{2\pi}\sigma_e)dx/\sqrt{2\pi}\sigma_g$

where $\gamma = \alpha/\sqrt{\sigma_g^2+\sigma_e^2}$. Note that for any non-zero average lifetime risk, $\gamma$ is fixed because $\alpha$ increases linearly with $\sqrt{\sigma_g^2+\sigma_e^2}$.

By definition heritability, $h = \sigma_G^2/(\sigma_G^2 + \sigma_E^2)$.

The integral within the brackets in the previous double integral can be expressed in terms of the error function, erf. Because the cumulative distribution function of normal distribution is given by $0.5(1 + erf(y/\sqrt{2}\sigma_e))$, the integral inside the brackets is $0.5 - 0.5erf([\gamma\sqrt{\sigma_g^2+\sigma_e^2}-x]/\sqrt{2}\sigma_e)$.

Thus, the probability that an individual is a true case and its genetic variable is greater than c can expressed as: $\int_{\beta\sigma_g}^{\infty} e^{-x^2/2\sigma_g^2}(0.5 - 0.5erf(\gamma f(h) - g(h)x/\sqrt{2}\sigma_g))dx/\sqrt{2\pi}\sigma_g$, where f(h) and g(h) are some functions of the heritability. Substituting $t = x/\sqrt{2}\sigma_g$ into this equation, we can see that $\sqrt{2}\sigma_g dt = dx$. Therefore, $P$ $(G > c$ and $G + E > \alpha)$ can be expressed as:

$$\int_{\beta/\sqrt{2}}^{\infty} e^{-t^2}(0.5 - 0.5erf(\gamma f(h) - g(h)t))dt/\sqrt{\pi}$$

Similarly, the probability that an individual is a true control and its genetic variable is greater than c i.e. $P(G > c$ and $G + E <= \alpha) = \int_{\beta/\sqrt{2}}^{\infty} e^{-t^2}(0.5 + 0.5erf(\gamma f(h) - g(h)t))dt/\sqrt{\pi}$.

Therefore, the true positive fraction for any given $\beta$ only depends on h and ALTR since: $TPF = P(G > c$ and $G + E > \alpha)/ALTR$. The same is also true for false positive fraction since $FPF = P(G > c$ and $G + E <= \alpha)/(1 - ALTR)$. Hence, the total area under the theoretical ROC curve, which is based on TPF and FPF at all possible values of $\beta$, is independent of $\sigma_e$ and $\sigma_g$.

## Proof for theoretical disease model 2

In this section, we prove a result similar to that in the previous section for disease model 2. In particular, we will show that if the relative risks of SNPs known to be associated with a disease and the risk-allele frequencies ($p_i$) are fixed, then the theoretical genetic maximum of the area under the ROC curve depends only on the heritability and the average lifetime risk of the disease. In model 2, the genetic variable is given by: $G = \sum \lambda_i X_i + G1$. Here $G1 \sim N(0, \sigma_{g1})$ and the $X_i$s are distributed according to a Binomial distribution of $B(2, p_i)$, where $p_i$ is the allele frequency of the risk allele at locus i. $B(2, p_i)$ gives the number of risk allele copies in an individual at locus i. $X_i = 0$ means homozygous for non-risk allele, $X_i = 1$ means heterozygote and $X_i = 2$ means homozygous for risk allele. The normal variable represents the unknown genetic component. As before, the environmental variable E is also normally distributed with mean 0 and standard deviation $\sigma_e$. The phenotype is given by $P = G + E$ and individuals with $P > \alpha$ are diseased whereas the rest are controls. $\alpha$ is chosen such that the fraction of diseased individuals equals the average lifetime risk of the disease.

Heritability for this model is $h = [\sigma_{g1}^2 + \sum 2\lambda_i^2 p_i(1-p_i)]/[\sigma_{g1}^2 + \sigma_e^2 + \sum 2\lambda_i^2 p_i(1-p_i)]$. Let us assume that the relative risks of the known SNPs for heterozygous genotypes are fixed and denote these by $RN_i$. By definition, the relative risk of heterozygote is given by:

$$RN_i = Pr(G + E > \alpha \mid X_i = 1)/Pr(G + E > \alpha \mid X_i = 0)$$
$$= [\sum Pr(G1 + E > \alpha - z - \lambda_i)Pr(W = z)]/$$
$$[\sum Pr(G1 + E > \alpha - z)P(W = z)],$$

where $W = \sum_{j \neq i} \lambda_j X_j$.

Let erf denote the error function and erfc denote the complementary error function (i.e. $1 - erf(x)$). Since G1+E is $N(0, \sqrt{\sigma_{g1}^2 + \sigma_e^2})$, the relative risk expressed in terms of complementary error function is given by: $\sum 0.5erfc[(\alpha - z - \lambda_i)/\sqrt{2(\sigma_{g1}^2+\sigma_e^2)}]Pr(W = z)/\sum 0.5erfc[(\alpha - z)/\sqrt{2(\sigma_{g1}^2+\sigma_e^2)}]Pr(W = z)$. Thus, if $\lambda_i$s with disease cutoff $\alpha$ represent the solutions for the SNPs for some choice of $\sqrt{\sigma_{g1}^2+\sigma_e^2}$, then $L\lambda_i$s with cutoff of $L\alpha$ will necessarily be solutions if the standard deviation of G1 and E are changed by a factor of L. This is because z is always a linear combination of $\lambda_i$s. Therefore, $\lambda_i/\sqrt{\sigma_{g1}^2+\sigma_e^2}$ and $\gamma = \alpha/\sqrt{\sigma_{g1}^2+\sigma_e^2}$ are independent of $\sqrt{\sigma_{g1}^2+\sigma_e^2}$ and depend on heritability and ALTR alone.

By definition, $h(\sigma_{g1}^2 + \sigma_e^2) = (1 - h)\sum 2\lambda_i^2 p_i(1-p_i) + \sigma_{g1}^2$. This therefore means that: $\sigma_{g1}^2/(\sigma_{g1}^2+\sigma_e^2) = h - (1-h)\sum 2\lambda_i^2 p_i(1-p_i)/(\sigma_{g1}^2+\sigma_e^2)$. Since $\lambda_i/\sqrt{(\sigma_{g1}^2+\sigma_e^2)}$ and $p_i$ are independent of $\sqrt{\sigma_{g1}^2+\sigma_e^2}$, $\sigma_{g1}^2/(\sigma_{g1}^2+\sigma_e^2)$ is a function of heritability and ALTR alone. Let $Z = \sum \lambda_i X_i$ and V denote the vector of $X_i$ values. Then, if $Z = z$ for $V = v$, $z/\sqrt{2}\sigma_{g1} = b(h, ALTR, v)$ is a function of the heritability, ALTR and v alone and is independent of $\sqrt{\sigma_{g1}^2+\sigma_e^2}$.

The true positive fraction is defined as: $Pr(G > c$ & $G + E > \alpha)/Pr(G + E > \alpha)$ where c denotes the cutoff for genetic variable. Let $\beta = c/\sigma_{g1}$. The numerator for TPF can be calculated as: $\sum Pr(V = v, Z = z)\int_{\beta\sigma_{g1}-z}^{\infty} e^{-x^2/2\sigma_{g1}^2}(\int_{\gamma\sqrt{\sigma_{g1}^2+\sigma_e^2}-x-z}^{\infty} e^{-y^2/2\sigma_e^2}dy/\sqrt{2\pi}\sigma_e)dx/\sqrt{2\pi}\sigma_{g1}$

Using the error function to express the cumulative distribution function of the normal distribution, $Pr(G > c$ & $G + E > \alpha)$ is:

$$\sum Pr(V = v, Z = z)\int_{\beta\sigma_{g1}-z}^{\infty} e^{-x^2/2\sigma_{g1}^2}(0.5 - 0.5erf[r(h, ALTR, v) - s(h, ALTR)x/\sqrt{2}\sigma_{g1}])dx/\sqrt{2\pi}\sigma_{g1}$$

where r and s are some functions. Substituting $t = x/\sqrt{2}\sigma_g$ into this equation, we can see that $\sqrt{2}\sigma_g dt = dx$. Therefore, $P(G > c$ and $G + E > \alpha)$ can be expressed as:

$$\sum Pr(V=v, Z=z) \int\limits_{(\beta/\sqrt{2}) - b(h,ALTR,v)}^{\infty} e^{-t^2}(0.5 - 0.5 \mathrm{erf}[r(h,ALTR,v) - s(h,ALTR)t])dt/\sqrt{\pi}$$

Similarly, the probability that an individual is a true control and its genetic variable is greater than c i.e.

$$P(G > c \text{ and } G + E <= \alpha) =$$

$$\sum Pr(V=v, Z=z) \int\limits_{(\beta/\sqrt{2}) - b(h,ALTR,v)}^{\infty} e^{-t^2}(0.5 + 0.5 \mathrm{erf}[r(h,ALTR,v) - s(h,ALTR)t])dt/\sqrt{\pi}$$

Note that $ALTR = P(G + E > \alpha)$ and $Pr(V=v, Z=z)$ is fixed if $p_i$s are fixed. Therefore, the true positive fraction for any given $\beta$ only depends on the h and ALTR. The same is also true for false positive fraction since $FPF = Pr(G > c \text{ and } G + E <= \alpha)/(1 - ALTR)$. So, the area under the theoretical ROC curve, which is based on TPF and FPF at all possible values of $\beta$, is independent of $\sigma_e$, $\sigma_{g1}$ and $\lambda_i$s.

## Solving for $\lambda_i / \sqrt{\sigma_{g1}^2 + \sigma_e^2}$

We first note that $1 - (\sigma_{g1}^2/(h(\sigma_{g1}^2 + \sigma_e^2))) = (1-h)\sum 2\lambda_i^2 p_i (1-p_i)/(h(\sigma_{g1}^2 + \sigma_e^2))$. So, this equation implies that $0 <= \lambda_i / \sqrt{\sigma_{g1}^2 + \sigma_e^2} <= \sqrt{h/(2p_i(1-p_i)(1-h))}$ since LHS is always less than 1. In practice, we can obtain a simultaneous solution for all $\lambda_i / \sqrt{\sigma_{g1}^2 + \sigma_e^2}$ by using the following iterative procedure:

Initially, determine the $\lambda_i / \sqrt{\sigma_{g1}^2 + \sigma_e^2}$ for each SNP assuming that it is the only SNP present (i.e. assuming $\lambda_j = 0$ for all j not equal to i using a binary search between 0 and $\sqrt{h/(2p_i(1-p_i)(1-h))}$(Note that $RN_i$ increases with $\lambda_i / \sqrt{\sigma_{g1}^2 + \sigma_e^2}$). These values will be our initial guesses for $\lambda_i / \sqrt{\sigma_{g1}^2 + \sigma_e^2}$. Then,

1) Determine $\lambda_1 / \sqrt{\sigma_{g1}^2 + \sigma_e^2}$ assuming that $\lambda_j / \sqrt{\sigma_{g1}^2 + \sigma_e^2}$ for other SNPs are equal to what was calculated in the previous step by using a binary search between 0 and

$$\sqrt{[(h/(1-h)) - \sum_{j \neq 1} 2\lambda_j^2 p_j(1-p_j)/(\sigma_{g1}^2 + \sigma_e^2)]/(2p_1(1-p_1))}.$$

2) Determine $\lambda_2 / \sqrt{\sigma_{g1}^2 + \sigma_e^2}$ assuming that $\lambda_j / \sqrt{\sigma_{g1}^2 + \sigma_e^2}$ for other SNPs are equal to what was calculated in the previous step by using a binary search between 0 and

$$\sqrt{[(h/(1-h)) - \sum_{j \neq 2} 2\lambda_j^2 p_j(1-p_j)/(\sigma_{g1}^2 + \sigma_e^2)]/(2p_2(1-p_2))}.$$

......

n) Determine $\lambda_n / \sqrt{\sigma_{g1}^2 + \sigma_e^2}$ assuming that $\lambda_j / \sqrt{\sigma_{g1}^2 + \sigma_e^2}$ for other SNPs are equal to what was calculated in the previous step by using a binary search between 0 and

$$\sqrt{[(h/(1-h)) - \sum_{j \neq n} 2\lambda_j^2 p_j(1-p_j)/(\sigma_{g1}^2 + \sigma_e^2)]/(2p_n(1-p_n))}.$$

If all $RN_j$ values are sufficiently close to the observed values, stop. Else go back to step 1.

Simulation experiments indicated that the above heuristic converges to a simultaneous solution for all $\lambda_j / \sqrt{\sigma_{g1}^2 + \sigma_e^2}$ whenever a solution exists.

## Obtaining robust estimates of heritability and lifetime risk from literature

Since heritability can vary by population, age, environmental variation, phenotypic definition, sample size or standard error; we sought multiple references and chose the most robust estimate based on the method of calculating heritability, sample size, ancestral origin and study population. If several articles had good methodology, we tried to pick one "in the middle" of the range of reported estimates. For lifetime risk, there is often not multiple references and sometimes we relied on incidence data.

## Obtaining lifetime risk estimates from Incidence data

When lifetime risk data was not available from the literature (for Crohn's disease and Rheumatoid Arthritis), we used incidence data to obtain an estimate of the average lifetime risk (ALTR) using a conversion formula. Namely, we used the following formula:$ALTR = (N/n_i) \times$ incidence, where N represents the total number of individuals in the US from the 1990 or 2000 census data depending on the study used to identify incidence, $n_i$ is the number of live births in the US in the year 2000; each from the appropriate gender and ethnicity. The main assumptions in this formula are: 1. Fixed population size. 2. Maximum life span for all.

We first validated our formula to determine if incidence data could incorrectly estimate lifetime risk using incidence and lifetime risk data from the Surveillance Epidemiology and End Results of the National Cancer Institute (USA) for a number of common cancers. Using our calculation with incidence data we estimated the published lifetime risk within 1% for breast, colon, prostate and lung cancers (results not shown). Thus, we are confident that our lifetime risk calculations are reliable.

## Author Contributions

Conceived and designed the experiments: BKP EH JW MC DS. Performed the experiments: BKP EH JW DJT. Analyzed the data: BKP EH JW DJT ES HT. Contributed reagents/materials/analysis tools: JW DJT ES HT. Wrote the paper: BKP EH DJT ES HT MC DS.